\documentclass[a4paper,10pt,twoside]{cpc-hepnp}

\usepackage{multicol}
\usepackage{graphicx}
\usepackage{booktabs}
\usepackage{amssymb,bm,mathrsfs,bbm,amscd}
\usepackage[tbtags]{amsmath}
\usepackage{lastpage}

\def\dab{\int^{\alpha_{max}}_{\alpha_{min}}d\alpha\int^{\beta_{max}}_{\beta_{min}}d\beta}
\def\qq{\langle\bar qq\rangle}
\def\GGa{\langle GG\rangle}
\def\GGb{\langle g_s^2GG\rangle}

\def\qGqa{\langle\bar qGq\rangle}
\def\qGqb{\langle\bar qg_s\sigma\cdot Gq\rangle}
\def\FF(s){\left[(\alpha+\beta)m_c^2-\alpha\beta s\right]}
\def\HH(s){\left[m_c^2-\alpha(1-\alpha) s\right]}
\def\non{\\ \nonumber}
\newcommand{\ket}{\rangle}
\newcommand{\bra}{\langle}

\begin{document}

\fancyhead[c]{\small Chinese Physics C~~~Vol. 37, No. 1 (2013)
010201} \fancyfoot[C]{\small 010201-\thepage}

\footnotetext[0]{Received 14 March 2009}

\title{Revisiting hidden-charm pentaquarks from QCD sum rules\thanks{Supported by National Natural Science Foundation of China (Grants No. 11722540 and No. 11261130311),
Fundamental Research Funds for the Central Universities, and Foundation for Young Talents in College of Anhui Province (Grants No. gxyq2018103).}}

\author{%
      Jia-Bing Xiang$^{1)}$
\quad Hua-Xing Chen$^{1;1)}$\email{hxchen@buaa.edu.cn}%
\quad Wei Chen$^{2;2)}$\email{chenwei29@mail.sysu.edu.cn}%
\quad Xiao-Bo Li$^{1}$
\\
\quad Xing-Qun Yao$^{1}$
\quad Shi-Lin Zhu$^{3,4,5;3)}$\email{zhusl@pku.edu.cn}%
}
\maketitle

\address{%
$^1$School of Physics and Beijing Key Laboratory of Advanced Nuclear Materials and Physics, Beihang University, Beijing 100191, China \\
$^2$School of Physics, Sun Yat-Sen University, Guangzhou 510275, China \\
$^3$School of Physics and State Key Laboratory of Nuclear Physics and Technology, Peking University, Beijing 100871, China \\
$^4$Collaborative Innovation Center of Quantum Matter, Beijing 100871, China \\
$^5$Center of High Energy Physics, Peking University, Beijing 100871, China \\
}

\begin{abstract}
We revisit the hidden-charm pentaquark states $P_c(4380)$ and $P_c(4450)$ using the method of QCD sum rules by requiring the pole contribution to be larger than or around 30\% to better insure the one-pole parametrization to be valid. We find two mixing currents and our results suggest that the $P_c(4380)$ and $P_c(4450)$ can be identified as hidden-charm pentaquark states having $J^P=3/2^-$ and $5/2^+$, respectively, while there still exist other possible spin-parity assignments, such as $J^P=3/2^+$ and $J^P=5/2^-$, which needs to be clarified in further theoretical and experimental studies.
\end{abstract}



\footnotetext[0]{\hspace*{-3mm}\raisebox{0.3ex}{$\scriptstyle\copyright$}2013
Chinese Physical Society and the Institute of High Energy Physics
of the Chinese Academy of Sciences and the Institute
of Modern Physics of the Chinese Academy of Sciences and IOP Publishing Ltd}%

\begin{multicols}{2}

\section{Introduction}
\label{sec:intro}

Many exotic hadrons have been discovered in the past decade due to significant experimental progresses~\cite{pdg},
such as the two hidden-charm pentaquark resonances $P_c(4380)$ and $P_c(4450)$ discovered by the LHCb Collaboration~\cite{lhcb1,lhcb2,lhcb3,lhcb4}.
Besides them, more exotic hadrons are likely to be observed in the future BaBar, Belle, BESIII, CMS and LHCb experiments, etc.
They are new blocks of QCD matter, providing important hints to deepen our understanding of the non-perturbative QCD, and
their relevant theoretical and experimental studies are opening a new page for the hadron physics~\cite{review1,review2,review3,review4,review5,review6}.

In the past year, the $P_c(4380)$ and $P_c(4450)$ have been studied by various methods and models in order to explain their nature.
There are many possible interpretations, such as meson-baryon molecules~\cite{molecules1,molecules2,molecules3,molecules4,molecules5,molecules6,molecules7,molecules8,molecules9,molecules10,molecules11,molecules12},
compact diquark-diquark-antiquark pentaquarks~\cite{diquark1,diquark2,diquark3,diquark4},
compact diquark-triquark pentaquarks~\cite{triquark1,triquark2},
genuine multiquark states other than molecules~\cite{pentaquark1,pentaquark2,pentaquark3,pentaquark4,pentaquark5,pentaquark6},
and kinematical effects related to thresholds and triangle singularity~\cite{triangle1,triangle2,triangle3,triangle4,triangle5}, etc.
Their productions and decay properties are also interesting~\cite{production1,production2,production3,production4,production5,production6,production7,production8,production9,production10,production11,production12,production13}.
More extensive discussions can be found in Refs.~\cite{others1,others2,others3}.

The preferred spin-parity assignments for the $P_c(4380)$ and $P_c(4450)$ states were suggested to be $(3/2^-, 5/2^+)$, while some other assignments were also suggested to be possible by the LHCb Collaboration, such as $(3/2^+, 5/2^-)$ and $(5/2^+, 3/2^-)$~\cite{lhcb1}. It is useful to study all these possible assignments theoretically in order to better understand their properties.

In this paper we shall use the method of QCD sum rule to study the possible spin-parity assignments of the $P_c(4380)$ and $P_c(4450)$. Before doing this, we shall reinvestigate our previous studies on the $P_c(4380)$ and $P_c(4450)$~\cite{chen1,chen2} by requiring the pole contribution to be larger than or around 30\% to better insure the one-pole parametrization to be valid, which value is just 10\% in our previous studies~\cite{chen1,chen2}.
We would like to note that there have been some experimental data on exotic hadrons, but they are not enough and more experimental results are necessary in order to make our theoretical analyses more reliable.

This paper is organized as follows: the above reinvestigation will be done in Sec.~2, numerical analyses will be done in Sec.~3, the investigation of hidden-charm pentaquark states of $J^P = 3/2^+$ and $J^P = 5/2^-$ will be done in Sec.~4, and the results will be discussed and summarized in Sec.~5.

\section{QCD sum rules analyses}
\label{sec:sumrule}

All the local hidden-charm pentaquark interpolating currents have been systematically constructed in Refs.~\cite{chen1,chen2}, and some of them were selected to perform QCD sum rule analyses. The results suggest that the $P_c(4380)$ and $P_c(4450)$ can be interpreted as hidden-charm pentaquark states composed of anti-charmed mesons and charmed baryons.
However, the analyses therein use one criterion which is not optimized, that is to require the pole contribution to be larger than 10\% to insure the one-pole parametrization to be valid. This value is not so significant, and accordingly, the question arises whether we can find a larger pole contribution to better insure the one-pole parametrization?

In the present study we try to answer this question in order to find better (more reliable) QCD sum rule results. Especially, we find the following two mixing currents:
\begin{eqnarray}
J_{\mu,3/2-} &=& \cos\theta_1 \times \xi_{36\mu} + \sin\theta_1 \times \psi_{9\mu}
\label{def:mix1}
\\ \nonumber &=& \cos\theta_1 \times [\epsilon^{abc} (u^T_a C \gamma_\nu \gamma_5 d_b) \gamma_\nu \gamma_5 c_c] [\bar c_d \gamma_\mu \gamma_5 u_d]
\\ \nonumber && + \sin\theta_1 \times [\epsilon^{abc} (u^T_a C \gamma_\nu u_b) \gamma_\nu \gamma_5 c_c] [\bar c_d \gamma_\mu d_d] \, ,
\\
J_{\mu\nu,5/2+} &=& \cos\theta_2 \times \xi_{15\mu\nu} + \sin\theta_2 \times \psi_{4\mu\nu}
\label{def:mix2}
\\ \nonumber &=& \cos\theta_2 \times [\epsilon^{abc} (u^T_a C \gamma_\mu \gamma_5 d_b) c_c] [\bar c_d \gamma_\nu u_d]
\\ \nonumber && + \sin\theta_2 \times [\epsilon^{abc} (u^T_a C \gamma_\mu u_b) c_c] [\bar c_d \gamma_\nu \gamma_5 d_d]
+ \{ \mu \leftrightarrow \nu \} \, ,
\end{eqnarray}
where $a \cdots d$ are color indices; $\theta_{1/2}$ are two mixing angles; $J_{\mu,3/2-}$ and $J_{\mu\nu,5/2+}$ have the spin-parity $J^P = 3/2^-$ and $5/2^+$, respectively.
The four single currents, $\xi_{36\mu}$, $\psi_{9\mu}$, $\xi_{15\mu\nu}$ and $\psi_{4\mu\nu}$, were first constructed in Refs.~\cite{chen1,chen2}. We can verify:
\begin{enumerate}

\item The current $\xi_{36\mu}$ well couples to the $S$-wave $[\Lambda_c(1P)\bar D_1]$, $P$-wave $[\Lambda_c(1P)\bar D]$, $P$-wave $[\Lambda_c\bar D_1]$, and $D$-wave $[\Lambda_c\bar D]$ channels, etc. Here the $\Lambda_c(1P)$ denotes the $\Lambda_c(2593)$ of $J^P = 1/2^-$ and $\Lambda_c(2625)$ of $J^P = 3/2^-$.

\item The current $\psi_{9\mu}$ well couples to the $S$-wave $[\Sigma_c \bar D^*]$ channel, etc.

\item The current $\xi_{15\mu\nu}$ well couples to the $S$-wave $[\Lambda_c(1P) \bar D^*]$ and $P$-wave $[\Lambda_c\bar D^*]$ channels, etc.

\item The current $\psi_{4\mu\nu}$ well couples to the $S$-wave $[\Sigma_c^* \bar D_1]$ and $P$-wave $[\Sigma_c^* \bar D]$ channels, etc.

\end{enumerate}

We shall use the above two mixing currents, $J_{\mu,3/2-}$ and $J_{\mu\nu,5/2+}$, to perform QCD sum rule analyses, and the results will be given in the next section. Before doing that we briefly introduce our approach here, and we refer interested readers to read Refs.~\cite{sumrule1,sumrule2,sumrule3,sumrule4,sumrule5,sumrule6} for details.

Firstly, we assume $J_{\mu,3/2-}$ and $J_{\mu\nu,5/2+}$ couple to physical states through
\begin{eqnarray}
\langle 0 | J_{\mu,3/2-} | X_{3/2-} \rangle &=& f_{X_{3/2-}} u_\mu (p) \, ,
\\ \langle 0 | J_{\mu\nu,5/2+} | X_{5/2+} \rangle &=& f_{X_{5/2+}} u_{\mu\nu} (p) \, ,
\end{eqnarray}
and write the two-point correlation functions as
\begin{eqnarray}
\label{pi:spin32} && \Pi_{\mu \nu,3/2-}\left(q^2\right)
\\ \nonumber &=& i \int d^4x e^{iq\cdot x} \langle 0 | T\left[J_{\mu,3/2-}(x) \bar J_{\nu,3/2-}(0)\right] | 0 \rangle
\\ \nonumber &=& \left(\frac{q_\mu q_\nu}{q^2}-g_{\mu\nu}\right) (q\!\!\!\slash~ + M_{X_{3/2-}}) \Pi_{3/2-}\left(q^2\right) + \cdots \, ,
\\ \label{pi:spin52} && \Pi_{\mu \nu \rho \sigma,5/2+}\left(q^2\right)
\\ \nonumber &=& i \int d^4x e^{iq\cdot x} \langle 0 | T\left[J_{\mu\nu,5/2+}(x) \bar J_{\rho\sigma,5/2+}(0)\right] | 0 \rangle
\\ \nonumber &=& \left(g_{\mu\rho}g_{\nu\sigma} + g_{\mu\sigma} g_{\nu\rho} \right) (q\!\!\!\slash~ + M_{X_{5/2+}}) \Pi_{5/2+}\left(q^2\right) + \cdots \, ,
\end{eqnarray}
where $\cdots$ contains non-relevant spin components.

We note that if the physical state has the opposite parity, the $\gamma_5$-coupling should be used~\cite{parity1,parity2,parity3,parity4}, for example, if
\begin{eqnarray}
\langle 0 | J_{\mu,3/2-} | X^\prime_{3/2+} \rangle &=& f_{X^\prime_{3/2+}} \gamma_5 u^\prime_\mu (p) \, ,
\end{eqnarray}
then
\begin{eqnarray}
&& \Pi_{\mu \nu,3/2+}\left(q^2\right)
\\ \nonumber &=& i \int d^4x e^{iq\cdot x} \langle 0 | T\left[J_{\mu,3/2-}(x) \bar J_{\nu,3/2-}(0)\right] | 0 \rangle
\\ \nonumber &=& \left(\frac{q_\mu q_\nu}{q^2}-g_{\mu\nu}\right) (q\!\!\!\slash~ - M_{X_{3/2+}}) \Pi_{3/2+}\left(q^2\right) + \cdots \, .
\end{eqnarray}
Hence, we can compare terms proportional to $\mathbf{1} \times g_{\mu\nu}$ and $q\!\!\!\slash \times g_{\mu\nu}$ to determine the parity of $X^{(\prime)}_{3/2\pm}$. Accordingly, in the present study we shall use the terms proportional to $\mathbf{1} \times g_{\mu\nu}$ and $\mathbf{1} \times g_{\mu\rho} g_{\nu\sigma}$ to evaluate masses of $X$'s, which are then compared with those proportional to $q\!\!\!\slash \times g_{\mu\nu}$ and $q\!\!\!\slash~ \times g_{\mu\rho} g_{\nu\sigma}$ to determine their parity.

At the hadron level, we use the dispersion relation to rewrite the two-point correlation function as
%
\begin{equation}
\label{eq:disper}
\Pi(q^2)={\frac{1}{\pi}}\int^\infty_{s_<}\frac{{\rm Im} \Pi(s)}{s-q^2-i\varepsilon}ds \, ,
\end{equation}
%
where $s_<$ is the physical threshold.
Its imaginary part is defined as the spectral function,
which can be evaluated by inserting intermediate hadron states $\sum_n|n\rangle\langle n|$,
but adopting the usual parametrization of one-pole dominance for the ground state $X$ together with a continuum contribution:
%
\begin{eqnarray}
\nonumber \rho(s) \equiv \frac{1}{\pi}{\rm Im}\Pi(s)
&=& \sum_n\delta(s-M^2_n)\langle 0|J|n\rangle\langle n|{\bar J}|0\rangle
\label{eq:rho}
\\ &=& f_X^2\delta(s-m_X^2)+ \mbox{continuum}\, .
\end{eqnarray}
%

At the quark and gluon level, we insert Eqs.~(\ref{def:mix1}--\ref{def:mix2}) into the two-point correlation functions (\ref{pi:spin32}--\ref{pi:spin52}), and calculate them using the method of operator product expansion (OPE).
In the present study we evaluate $\rho(s)$ at the leading order on $\alpha_s$ and up to dimension eight. To do this we have calculated the perturbative term, the quark condensate $\langle \bar q q \rangle$, the gluon condensate $\langle g_s^2 GG \rangle$, the quark-gluon condensate $\langle g_s \bar q \sigma G q \rangle$, and their combinations $\langle \bar q q \rangle^2$ and $\langle \bar q q \rangle\langle g_s \bar q \sigma G q \rangle$.
We find that the $D=4$ term $m_c \langle \bar q q \rangle$ and the $D=6$ term $m_c \langle g_s \bar q \sigma G q \rangle$ are important power corrections to the correlation functions.
Note that we have assumed the vacuum saturation for higher dimensional operators such as $\bra 0 | \bar q q \bar q q |0\ket \sim \bra 0 | \bar q q |0\ket \bra 0|\bar q q |0\ket$, and this can lead to some systematic uncertainties.

Finally, we perform the Borel transform at both the hadron and quark-gluon levels, and express the two-point correlation function as
%
\begin{equation}
\label{eq:borel}
\Pi^{(all)}(M_B^2)\equiv\mathcal{B}_{M_B^2}\Pi(p^2) = \int^\infty_{s_<} e^{-s/M_B^2} \rho(s) ds \, .
\end{equation}
%
After assuming that the continuum contribution can be well approximated by the OPE spectral density above a threshold value $s_0$,
we obtain the sum rule relation
%
\begin{eqnarray}
\label{eq:mass}
M^2_X(s_0, M_B) &=& {\int^{s_0}_{s_<} e^{-s/M_B^2} \rho(s) s ds \over \int^{s_0}_{s_<} e^{-s/M_B^2} \rho(s) ds} \, .
\end{eqnarray}
%

We use the mixing current $J_{\mu,3/2-}$ defined in Eq.~(\ref{def:mix1}) to perform sum rule analyses, and the terms proportional to $\mathbf{1} \times g_{\mu\nu}$ are shown in Eq.~(\ref{ope:mix1}), where $t_1 = \cos\theta_1$ and $t_2 = \sin\theta_1$. Those proportional to $q\!\!\!\slash~\times g_{\mu\nu}$ are listed in Eq.~(\ref{ope:mix2}), which are almost the same as the former ones, suggesting that the state coupled by $J_{\mu,3/2-}$ has the spin-parity $J^P = 3/2^-$. Similarly, we use $J_{\mu\nu,5/2+}$ defined in Eq.~(\ref{def:mix2}) to perform sum rule analyses, and the terms proportional to $\mathbf{1} \times g_{\mu\nu}$ and $q\!\!\!\slash~\times g_{\mu\nu}$ are listed in Eqs.~(\ref{ope:mix3}) and (\ref{ope:mix4}), respectively. We find its relevant state has the spin-parity $J^P = 5/2^+$. These two sum rules will be used to perform numerical analyses in the next section.

\newcounter{mytempeqncnt1}
\begin{figure*}[!hbt]
\small
\hrulefill
\begin{eqnarray}
\rho_{3/2-,1}(s) &=& \rho^{pert}_{3/2-,1}(s) + \rho^{\qq}_{3/2-,1}(s) + \rho^{\GGa}_{3/2-,1}(s) + \rho^{\qq^2}_{3/2-,1}(s) + \rho^{\qGqa}_{3/2-,1}(s) + \rho^{\qq\qGqa}_{3/2-,1}(s) \, ,
\label{ope:mix1}
\non
\rho^{pert}_{3/2-,1}(s) &=& \frac{m_c}{3932160\pi^8}\dab \Bigg\{ \FF(s)^5 \times \frac{(1 - \alpha - \beta)^3 (\alpha + \beta + 3) \left(11 t_1^2-4 t_1 t_2+24 t_2^2\right)}{\alpha^5\beta^4} \Bigg\} \, ,
\non
\rho^{\qq}_{3/2-,1}(s) &=& \frac{m_c^2\qq}{3072\pi^6} \dab \Bigg\{ \FF(s)^3 \times \frac{(1 - \alpha - \beta)^2 \left(3 t_1^2-2 t_1 t_2-6 t_2^2\right)}{\alpha^3\beta^3} \Bigg\} \, ,
\non
\rho^{\GGa}_{3/2-,1}(s) &=& - \frac{m_c\GGb}{28311552 \pi^8}\dab\Bigg\{
\FF(s)^3 \times \Big(
\frac{432(1 - \alpha - \beta) (\alpha + \beta + 1) \left(t_1^2 + 2t_2^2\right)}{\alpha^3\beta^2}
\non
&& - \frac{12 (1 - \alpha - \beta)^2 (\alpha +\beta -4) \left(7 t_1^2-4 t_1t_2+16t_2^2\right)}{\alpha^3\beta^3}
- \frac{36 (1 - \alpha - \beta)^2 (\alpha +\beta +2) \left(7 t_1^2-4 t_1 t_2+16 t_2^2\right)}{\alpha^4\beta^2}
\non
&& + \frac{(1 - \alpha - \beta)^3  (\alpha +\beta -5) (t_1^2 +4t_1t_2)}{\alpha^4\beta^3}
- \frac{6 (1- \alpha - \beta)^3 (\alpha +\beta +3) \left(11 t_1^2-4 t_1 t_2+24 t_2^2\right)}{\alpha^5\beta^2}\Big)
\non
&& - 6 m_c^2 \FF(s)^2 \times \left(11 t_1^2-4 t_1 t_2+24 t_2^2\right)
\times \Big( \frac{(1- \alpha - \beta)^3 (\alpha +\beta +3) }{\alpha^2\beta^4}
+ \frac{(1- \alpha - \beta)^3 (\alpha +\beta +3)}{\alpha^5\beta} \Big) \Bigg\} \, ,
\non
\rho^{\qGqa}_{3/2-,1}(s) &=& - \frac{m_c^2\qGqb}{8192\pi^6} \dab \Bigg\{ \FF(s)^2
\non
&& \times \Big( \frac{(1- \alpha - \beta) \left(13 t_1^2-8 t_1 t_2-24 t_2^2\right)}{\alpha^2\beta^2}
+ \frac{2(1- \alpha - \beta)^2 t_1 t_2 }{\alpha^3\beta^2} \Big) \Bigg\} \, ,
\non
\rho^{\qq^2}_{3/2-,1}(s)&=& \frac{m_c\qq^2}{1536\pi^4} \dab \Bigg\{ \FF(s)^2 \times \frac{(\alpha +\beta ) \left(21 t_1^2-4 t_1t_2-48 t_2^2\right)}{\alpha^2\beta} \Bigg\} \, ,
\non
\rho^{\qq\qGqa}_{3/2-,1}(s)&=& \frac{m_c\qq\qGqb}{9216\pi^4} \dab \Bigg\{ \FF(s)
\non
&&
\times\Big( \frac{3 \left(47 t_1^2 -12 t_1t_2 -96 t_2^2\right)}{\alpha}
- \frac{2 (\alpha +\beta -2) \left(t_1^2+4t_1 t_2\right)}{\alpha\beta}
+ \frac{3 (\alpha +\beta ) \left(21 t_1^2-4 t_1 t_2-48 t_2^2\right)}{\alpha^2}
\Big) \Bigg\}
\non
&& - \frac{m_c\qq\qGqb}{3072\pi^4} \int^{\alpha_{max}}_{\alpha_{min}}d\alpha \Bigg\{ \HH(s) \times \frac{47 t_1^2-12t_1 t_2-96 t_2^2}{\alpha} \Bigg\} \, .
\end{eqnarray}
\hrulefill
\vspace*{4pt}
\end{figure*}

\newcounter{mytempeqncnt2}
\begin{figure*}[!hbt]
\small
\hrulefill
\begin{eqnarray}
\rho_{3/2-,2}(s) &=& \rho^{pert}_{3/2-,2}(s) + \rho^{\qq}_{3/2-,2}(s) + \rho^{\GGa}_{3/2-,2}(s) + \rho^{\qq^2}_{3/2-,2}(s) + \rho^{\qGqa}_{3/2-,2}(s) + \rho^{\qq\qGqa}_{3/2-,2}(s) \, ,
\label{ope:mix2}
\non
\rho^{pert}_{3/2-,2}(s) &=& \frac{1}{3932160\pi^8}\dab \Bigg\{ \FF(s)^5
\non
&& \times\Big( \frac{8 (1 - \alpha - \beta)^3 \left(11 t_1^2 -4 t_1 t_2 +24 t_2^2\right)}{\alpha^4\beta^4}
-\frac{3 (1 - \alpha - \beta)^4 \left( 7 t_1^2 -4 t_1 t_2 +16 t_2^2 \right)}{\alpha^4\beta^4}\Big) \Bigg\} \, ,
\non
\rho^{\qq}_{3/2-,2}(s) &=& \frac{mc \qq}{12288 \pi^6} \dab \Bigg\{ \FF(s)^3 \times \frac{(1 -\alpha -\beta)^2 \left(21 t_1^2 -4 t_1 t_2 -48 t_2^2 \right)}{\alpha^2\beta^3} \Bigg\} \, ,
\non
\rho^{\GGa}_{3/2-,2}(s) &=& - \frac{\GGa}{28311552 \pi^8} \dab \Bigg\{
\FF(s)^3 \times \Big( \frac{144 (1 -\alpha -\beta) (\alpha +\beta +1) \left( t_1^2 + 4 t_1 t_2 \right) }{\alpha^2\beta^2}
\non
&& + \frac{96 (1 -\alpha -\beta)^2 \left( 5 t_1^2 -4 t_1 t_2 +12 t_2^2\right)}{\alpha^2\beta^3}
- \frac{108 (1 -\alpha -\beta)^2 \left( 7 t_1^2 -4 t_1 t_2 +16 t_2^2\right)}{\alpha^3\beta^2}
+ \frac{4 (1 -\alpha -\beta)^3 \left( 35 t_1^2 -52 t_1 t_2 +96 t_2^2\right)}{\alpha^2\beta^3}
\non
&&+ \frac{4 (1 -\alpha -\beta)^3 \left( 61 t_1^2 -44 t_1 t_2 +144 t_2^2\right)}{\alpha^3\beta^2}
+ \frac{ (1 -\alpha -\beta)^3 (\alpha +\beta -5)\left( t_1^2 +4 t_1 t_2\right)}{\alpha^3\beta^2}\Big)
\non
&&-6 mc^2 \FF(s)^2 \times \Big( \frac{8 (1 -\alpha -\beta)^3 \left( 11 t_1^2 -4 t_1 t_2 +24 t_2^2\right)}{\alpha\beta^4}
+ \frac{8 (1 -\alpha -\beta)^3 \left( 11 t_1^2 -4 t_1 t_2 +24 t_2^2\right)}{\alpha^4\beta}
\non
&&- \frac{3 (1 -\alpha -\beta)^4 \left( 7 t_1^2 -4 t_1 t_2 +16 t_2^2\right)}{\alpha\beta^4}
- \frac{3 (1 -\alpha -\beta)^4 \left( 7 t_1^2 -4 t_1 t_2 +16 t_2^2\right)}{\alpha^4\beta}\Big)\Bigg\}\, ,
\non
\rho^{\qGqa}_{3/2-,2}(s) &=& -\frac{mc \qGqa}{32768 \pi^6} \dab \Bigg\{ \FF(s)^2
\non
&&\times\Big( \frac{8 (1 -\alpha -\beta) \left( 13 t_1^2 -4 t_1 t_2 -24 t_2^2\right)}{\alpha\beta^2}
+ \frac{(1 -\alpha -\beta)^2 \left(t_1^2 +4 t_1 t_2\right)}{\alpha^2\beta^2}\Big)\Bigg\}\, ,
\non
\rho^{\qq^2}_{3/2-,2}(s) &=& \frac{\qq^2}{1536 \pi^4} \dab \Bigg\{ \FF(s)^2 \times \Big( \frac{4 \left( 3 t_1^2 -2 t_1 t_2 -6 t_2^2\right)}{\alpha\beta} - \frac{(1 -\alpha -\beta) \left( 11 t_1^2 -4 t_1 t_2 -24 t_2^2\right)}{\alpha\beta} \Big)\Bigg\}\, ,
\non
\rho^{\qq\qGqa}_{3/2-,2}(s) &=& \frac{\qq\qGqb}{9216\pi^4} \dab \Bigg\{ \FF(s) \times
\Big( 4 \left(18 t_1^2 -7 t_1t_2 -36 t_2^2 \right)
+ \frac{3 \left( 11 t_1^2 -4 t_1 t_2 -24 t_2^2\right)}{\alpha}
\non
&&- \frac{(3 t_1^2 -16t_1 t_2)}{\beta}
- \frac{ (1 -\alpha -\beta)\left( 31 t_1^2 -12 t_1 t_2 -72 t_2^2\right)}{\alpha}
- \frac{(1- \alpha -\beta ) \left(t_1^2 -8 t_1 t_2\right)}{\beta}
\Big) \Bigg\}
\non
&& - \frac{\qq\qGqb}{3072\pi^4} \int^{\alpha_{max}}_{\alpha_{min}}d\alpha \Bigg\{ \HH(s) \times \Big(25 t_1^2 -16t_1 t_2 -48 t_2^2\Big) \Bigg\} \, .
\end{eqnarray}
\hrulefill
\vspace*{4pt}
\end{figure*}

\newcounter{mytempeqncnt3}
\begin{figure*}[!hbt]
\small
\hrulefill
\begin{eqnarray}
\rho_{5/2+,1}(s) &=& \rho^{pert}_{5/2+,1}(s) + \rho^{\qq}_{5/2+,1}(s) + \rho^{\GGa}_{5/2+,1}(s) + \rho^{\qq^2}_{5/2+,1}(s) + \rho^{\qGqa}_{5/2+,1}(s) + \rho^{\qq\qGqa}_{5/2+,1}(s) \, ,
\label{ope:mix3}
\non
\rho^{pert}_{5/2+,1}(s) &=& -\frac{m_c}{4915200\pi^8}\dab \Bigg\{ \FF(s)^5 \times \left(5 t_1^2 -4 t_1 t_2 +12 t_2^2\right)
\non
&& \times \Big( \frac{10 (1 - \alpha - \beta)^3 (\alpha +\beta +1) }{\alpha^5\beta^4}
-\frac{(1 - \alpha - \beta)^4 (\alpha +\beta +4) }{\alpha^5\beta^4}\Big) \Bigg\} \, ,
\non
\rho^{\qq}_{5/2+,1}(s) &=& \frac{mc^2 \qq}{18432 \pi^6} \dab \Bigg\{ \FF(s)^3 \times \frac{(1 -\alpha -\beta)^2 (\alpha +\beta +2) \left(t_1 -2 t_1\right) \left(5 t_1 +6 t_2\right)}{\alpha^3\beta^3} \Bigg\} \, ,
\non
\rho^{\GGa}_{5/2+,1}(s) &=& \frac{m_c \GGa}{35389440 \pi^8} \dab \Bigg\{
\FF(s)^3 \times \Big( \frac{360 (1 -\alpha -\beta) \left(t_1^2 +2 t_2^2 \right) }{\alpha^3\beta^2}
\non
&& - \frac{120 (1 -\alpha -\beta)^2 \left( t_1^2 +4 t_1 t_2 \right)}{\alpha^3\beta^2}
+ \frac{90 (1 -\alpha -\beta)^2 \left( 3 t_1^2 -4 t_1 t_2 +8 t_2^2\right)}{\alpha^3\beta^3}
- \frac{40 (1 -\alpha -\beta)^3 \left( t_1^2 -8 t_1 t_2 +6 t_2^2\right)}{\alpha^3\beta^2}
\non
&&- \frac{20 (1 -\alpha -\beta)^3 \left( t_1^2 -4 t_1 t_2 \right)}{\alpha^3\beta^3}
- \frac{60 (1 -\alpha -\beta)^3 \left( 5 t_1^2 -4 t_1 t_2 +12 t_2^2\right )}{\alpha^5\beta^2}
- \frac{5 (1 -\alpha -\beta)^4 \left( 7 t_1^2 -20 t_1 t_2 +24 t_2^2\right )}{\alpha^3\beta^3}
\non
&& + \frac{6 (1 -\alpha -\beta)^4 (\alpha +\beta +4) \left( 5 t_1^2 -4 t_1 t_2 +12 t_2^2\right )}{\alpha^5\beta^2}\Big)
\non
&& - 6 mc^2 \FF(s)^2 \times \left( 5 t_1^2 -4 t_1 t_2 +12 t_2^2 \right)
\non
&& \times \Big( \frac{10 (1 -\alpha -\beta)^3 }{\alpha^2\beta^4}
+ \frac{10 (1 -\alpha -\beta)^3 }{\alpha^5\beta}
- \frac{ (1 -\alpha -\beta)^4 (\alpha +\beta +4) }{\alpha^2\beta^4}
- \frac{ (1 -\alpha -\beta)^4 (\alpha +\beta +4) }{\alpha^5\beta}\Big)\Bigg\}\, ,
\non
\rho^{\qGqa}_{5/2+,1}(s) &=& -\frac{mc^2 \qGqa}{24576 \pi^6} \dab \Bigg\{ \FF(s)^2 \times
\frac{(1 -\alpha -\beta) (\alpha +\beta +1) \left( t_1 -2 t_1 \right) \left( 17 t_1 +18 t_2 \right)}{\alpha^2\beta^2} \Bigg\}\, ,
\non
\rho^{\qq^2}_{5/2+,1}(s) &=& -\frac{mc \qq^2}{768 \pi^4} \dab \Bigg\{ \FF(s)^2 \times
\frac{(\alpha +\beta) \left( 5 t_1^2 -4 t_1 t_2 -12 t_2^2 \right)}{\alpha^2\beta} \Bigg\}\, ,
\non
\rho^{\qq\qGqa}_{5/2+,1}(s) &=& - \frac{m_c \qq\qGqb}{4608\pi^4} \dab \Bigg\{ \FF(s)
\non
&&\times\Big( \frac{4 \left(t_1 -2 t_1\right) \left(8 t_1 +9 t_2\right)}{\alpha}
- \frac{(\alpha +\beta ) \left(t_1^2 -8 t_1 t_2\right)}{\alpha\beta}\Big) \Bigg\}
\non
&& + \frac{m_c \qq\qGqb}{1152\pi^4} \int^{\alpha_{max}}_{\alpha_{min}}d\alpha \Bigg\{ \HH(s) \times \frac{\left(t_1 -2 t_1\right) \left(8 t_1 +9 t_2\right)}{\alpha} \Bigg\} \, .
\end{eqnarray}
\hrulefill
\vspace*{4pt}
\end{figure*}

\newcounter{mytempeqncnt4}
\begin{figure*}[!hbt]
\small
\hrulefill
\begin{eqnarray}
\rho_{5/2+,2}(s) &=& \rho^{pert}_{5/2+,2}(s) + \rho^{\qq}_{5/2+,2}(s) + \rho^{\GGa}_{5/2+,2}(s) + \rho^{\qq^2}_{5/2+,2}(s) + \rho^{\qGqa}_{5/2+,2}(s) + \rho^{\qq\qGqa}_{5/2+,2}(s) \, ,
\label{ope:mix4}
\non
\rho^{pert}_{5/2+,2}(s) &=& -\frac{1}{4915200\pi^8}\dab \Bigg\{ \FF(s)^5 \times \left(5 t_1^2 -4 t_1 t_2 +12 t_2^2\right)
\non
&& \times\Big( \frac{10 (1 - \alpha - \beta)^3 }{\alpha^4\beta^4}
-\frac{(1 - \alpha - \beta)^4 (\alpha +\beta +4) }{\alpha^4\beta^4}\Big) \Bigg\} \, ,
\non
\rho^{\qq}_{5/2+,2}(s) &=& \frac{mc \qq}{18432 \pi^6} \dab \Bigg\{ \FF(s)^3 \times \frac{(1 -\alpha -\beta)^2 (\alpha +\beta +2) \left(t_1 -2 t_1\right) \left(5 t_1 +6 t_2\right)}{\alpha^2\beta^3} \Bigg\} \, ,
\non
\rho^{\GGa}_{5/2+,2}(s) &=&  \frac{\GGa}{35389440 \pi^8} \dab \Bigg\{
5 \FF(s)^3 \times \Big( \frac{72 (1 -\alpha -\beta) \left(t_1^2 +2 t_2^2 \right) }{\alpha^2\beta^2}
\non
&& - \frac{24 (1 -\alpha -\beta)^2 \left( t_1^2 +4 t_1 t_2 \right)}{\alpha^2\beta^2}
+ \frac{18 (1 -\alpha -\beta)^2 \left( 3 t_1^2 -4 t_1 t_2 +8 t_2^2\right)}{\alpha^2\beta^3}
- \frac{8 (1 -\alpha -\beta)^3 \left( t_1^2 -8 t_1 t_2 +6 t_2^2\right)}{\alpha^2\beta^2}
\non
&&- \frac{4 (1 -\alpha -\beta)^3 \left( t_1^2 +4 t_1 t_2 \right)}{\alpha^2\beta^3}
- \frac{ (1 -\alpha -\beta)^4 \left( 7 t_1^2 -20 t_1 t_2 +24 t_2^2 \right)}{\alpha^2\beta^3}\Big)
\non
&& - 6 mc^2 \FF(s)^2 \times \left( 5 t_1^2 -4 t_1 t_2 +12 t_2^2\right)
\non
&&\times \Big( \frac{10 (1 -\alpha -\beta)^3 }{\alpha\beta^4}
+ \frac{10 (1 -\alpha -\beta)^3}{\alpha^4\beta}
- \frac{ (1 -\alpha -\beta)^4 (\alpha +\beta +4) }{\alpha\beta^4}
- \frac{ (1 -\alpha -\beta)^4 (\alpha +\beta +4) }{\alpha^4\beta}\Big)\Bigg\}\, ,
\non
\rho^{\qGqa}_{5/2+,2}(s) &=& -\frac{mc \qGqa}{24576 \pi^6} \dab \Bigg\{ \FF(s)^2 \times
\frac{(1 -\alpha -\beta) (\alpha +\beta +1) \left( t_1 -2 t_1 \right) \left( 17 t_1 +18 t_2 \right)}{\alpha\beta^2} \Bigg\}\, ,
\non
\rho^{\qq^2}_{5/2+,2}(s) &=& -\frac{\qq^2}{768 \pi^4} \dab \Bigg\{ \FF(s)^2 \times
\frac{(\alpha +\beta) \left( 5 t_1^2 -4 t_1 t_2 -12 t_2^2 \right)}{\alpha\beta} \Bigg\}\, ,
\non
\rho^{\qq\qGqa}_{5/2+,2}(s) &=& - \frac{\qq\qGqb}{4608\pi^4} \dab \Bigg\{ \FF(s) \times
\Big( 4 \left(t_1 -2 t_1\right) \left(8 t_1 +9 t_2\right)
- \frac{(\alpha +\beta ) \left(t_1^2 -8 t_1 t_2\right)}{\beta}\Big) \Bigg\}
\non
&& + \frac{\qq\qGqb}{1152\pi^4} \int^{\alpha_{max}}_{\alpha_{min}}d\alpha \Bigg\{ \HH(s) \times \left(t_1 -2 t_1\right) \left(8 t_1 +9 t_2\right)\Bigg\} \, .
\end{eqnarray}
\hrulefill
\vspace*{4pt}
\end{figure*}

\section{Numerical Analyses}
\label{sec:numerical}

In this section we use the sum rules for $J_{\mu,3/2-}$ and $J_{\mu\nu,5/2+}$ to perform numerical analyses.
Various condensates inside these equations take the following values~\cite{pdg,parameter1,parameter2,parameter3,parameter4,parameter5,parameter6,parameter7,parameter8}:
\begin{eqnarray}
\nonumber && \langle \bar qq \rangle = - (0.24 \pm 0.01)^3 \mbox{ GeV}^3 \, ,
\\ && \langle g_s^2GG\rangle =(0.48 \pm 0.14) \mbox{ GeV}^4\, ,
\\ \nonumber && \langle g_s \bar q \sigma G q \rangle = M_0^2 \times \langle \bar qq \rangle\, ,
\\ \nonumber && M_0^2 = - 0.8 \mbox{ GeV}^2\, .
\end{eqnarray}
We also need the charm and bottom quark masses, for which we use the running mass in the $\overline{MS}$ scheme~\cite{pdg,parameter1,parameter2,parameter3,parameter4,parameter5,parameter6,parameter7,parameter8}:
\begin{eqnarray}
&& m_c = 1.275 \pm 0.025 \mbox{ GeV} \, ,
\\ \nonumber && m_b = 4.18^{+0.04}_{-0.03} \mbox{ GeV} \, .
\end{eqnarray}

There are altogether three free parameters in Eq.~(\ref{eq:mass}): the mixing angles $\theta_{1/2}$, the Borel mass $M_B$, and the threshold value $s_0$. We find that after fine-tuning the two mixing angles to be $\theta_1=-42^\circ$ and $\theta_2=-45^\circ$, the following three criteria can be satisfied so that reliable sum rule results can be achieved:
\begin{enumerate}

\item  The first criterion is used to insure the convergence of the OPE series, i.e., we require the dimension eight to be less than 10\%, which can be used to determine the lower limit of the Borel mass:
%
\begin{eqnarray}
\mbox{CVG} &\equiv& \left|\frac{ \Pi_{\langle \bar q q \rangle\langle g_s \bar q \sigma G q \rangle}(\infty, M_B) }{ \Pi(\infty, M_B) }\right| \leq 10\% \, .
\label{eq_convergence}
\end{eqnarray}
%

\item The second criterion is used to insure the one-pole parametrization to be valid, i.e., we require the pole contribution (PC) to be larger than or around 30\%, which can be used to determine the upper limit of the Borel mass:
%
\begin{eqnarray}
\mbox{PC}(s_0, M_B) \equiv \frac{ \Pi(s_0, M_B) }{ \Pi(\infty, M_B) } \gtrsim 30\% \, .
\label{eq_pole}
\end{eqnarray}
%
This criterion better insure the one-pole parametrization than that used in Refs.~\cite{chen1,chen2} which only requires PC$ \geq 10\%$.

\item The third criterion is to require that both the $s_0$ and the $M_B$ dependence of the mass prediction be the weakest in order to obtain reliable mass predictions.

\end{enumerate}

\begin{figure*}[hbt]
\begin{center}
\scalebox{0.4}{\includegraphics{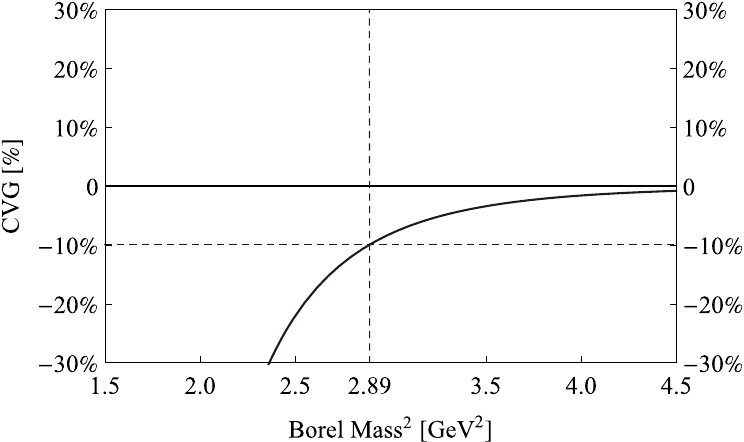}}
\scalebox{0.4}{\includegraphics{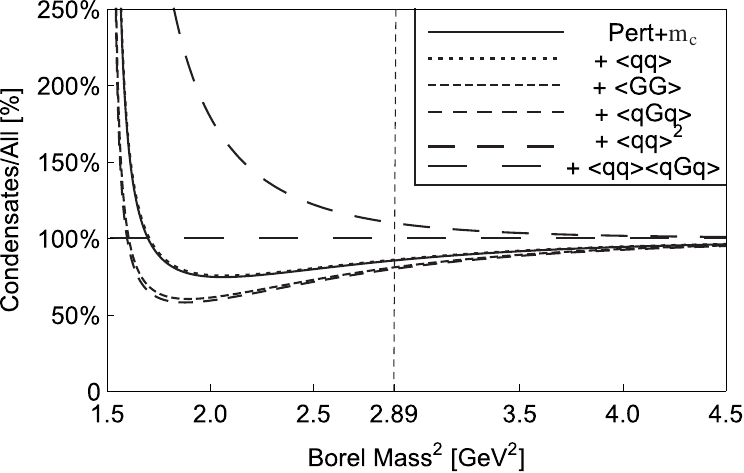}}
\scalebox{0.4}{\includegraphics{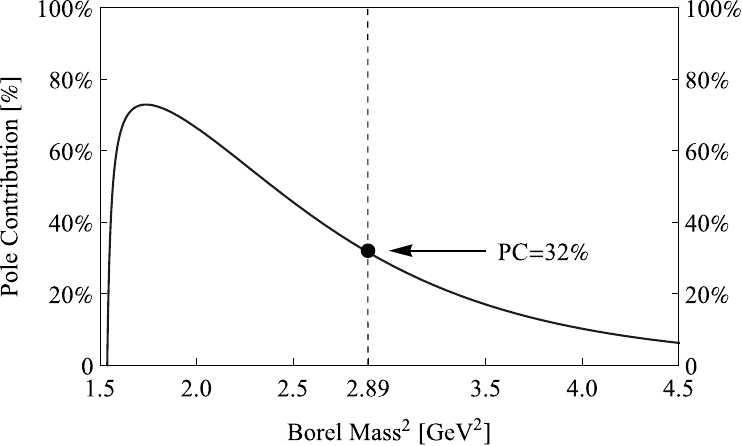}}
\caption{In the left panel we show CVG, defined in Eq.~(\ref{eq_convergence}), as a function of the Borel mass $M_B$.
In the middle panel we show the relative contribution of each term on the OPE expansion, as a function of the Borel mass $M_B$.
In the right panel we show the variation of PC, defined in Eq.~(\ref{eq_pole}), as a function of the Borel mass $M_B$.
Here we use the current $J_{\mu,3/2-}$ of $J^P = 3/2^-$, and choose $\theta_1=-42^\circ$ and $s_0 = 23$ GeV$^2$.}
\label{fig:cvg1}
\end{center}
\end{figure*}

\begin{figure*}[!hbt]
\begin{center}
\scalebox{0.4}{\includegraphics{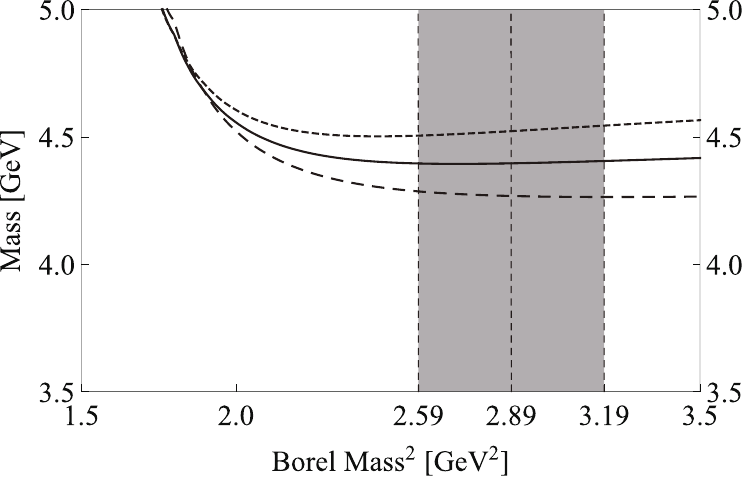}}
\scalebox{0.4}{\includegraphics{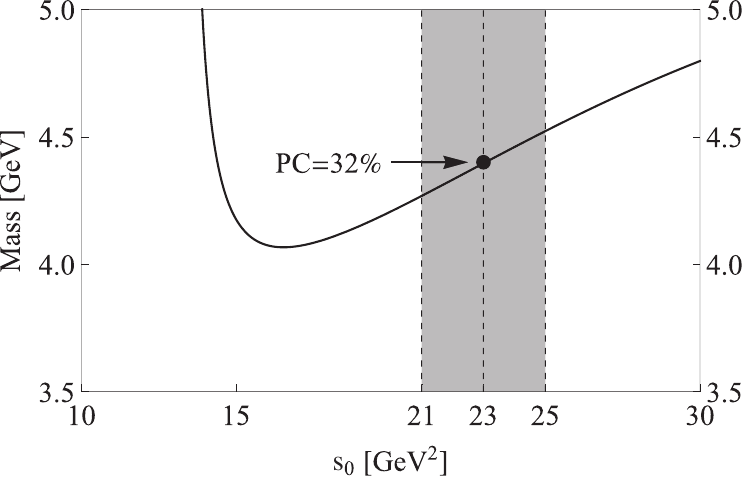}}
\scalebox{0.4}{\includegraphics{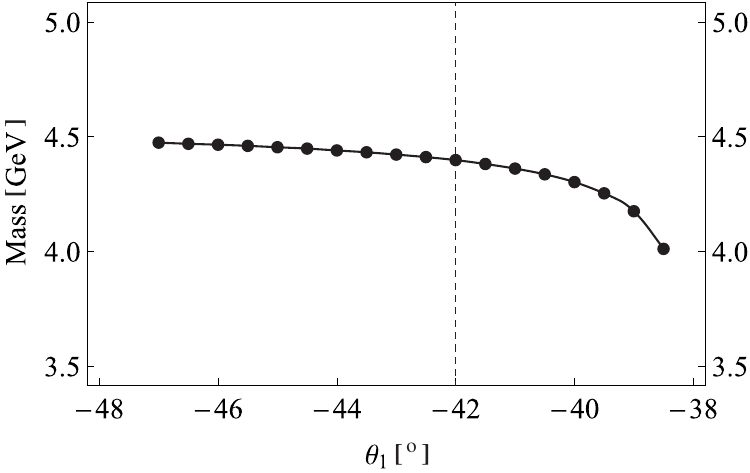}}
\caption{Variations of $M_{3/2^-}$ with respect to the Borel mass $M_B$ (left), the threshold value $s_0$ (middle) and the mixing angle $\theta_1$ (right),
calculated using the current $J_{\mu,3/2-}$ of $J^P = 3/2^-$.
In the left figure, the long-dashed, solid and short-dashed curves are obtained with $\theta_1=-42^\circ$ and for $s_0 = 21$, $23$ and $25$ GeV$^2$, respectively.
In the middle figure, the curve is obtained with $\theta_1=-42^\circ$ and $M_B^2 = 2.89$ GeV$^2$.
In the right figure, the curve is obtained for $s_0 = 23$ GeV$^2$ and with $M_B$ satisfying CVG$=10\%$.}
\label{fig:mix1mb}
\end{center}
\end{figure*}

We use the sum rules (\ref{ope:mix1}) for the current $J_{\mu,3/2-}$ as an example.
Firstly, we fix $\theta_1=-42^\circ$ and $s_0 = 23$ GeV$^2$, and show CVG as a function of $M_B$ in the left panel of Fig.~\ref{fig:cvg1}. We find that the OPE convergence improves with the increase of $M_B$, and the first criterion requires that $M_B^2 \geq 2.89$ GeV$^2$. We also show the relative contribution of each term in the middle panel of Fig.~\ref{fig:cvg1}, again we find that a good convergence can be achieved in the same region $M_B^2 \geq 2.89$ GeV$^2$.
Secondly, we still fix $\theta_1=-42^\circ$ and $s_0 = 23$ GeV$^2$, and show PC as a function of $M_B$ in the right panel of Fig.~\ref{fig:cvg1}. We find that PC decreases with the increase of $M_B$, and PC = 32\% when $M_B^2 = 2.89$ GeV$^2$. Accordingly, we fix the Borel mass to be $M_B^2 = 2.89$ GeV$^2$ and choose $2.59$ GeV$^2< M_B^2 < 3.19$ GeV$^2$ as our working region. We show variations of $M_X$ with respect to $M_B$ in the left panel of Fig.~\ref{fig:mix1mb}, and find that the mass curves are quite stable around $M_B^2 = 2.89$ GeV$^2$ as well as inside the Borel window $2.59$ GeV$^2< M_B^2 < 3.19$ GeV$^2$.

To use the third criterion to determine $s_0$, we show variations of $M_X$ with respect to $s_0$ in the middle panel of Fig.~\ref{fig:mix1mb} when fixing $\theta_1=-42^\circ$.
The mass curves have a minimum against $s_0$ when $s_0$ is around 17 GeV$^2$, so the $s_0$ dependence of the mass prediction is the weakest at this point. However, the pole contribution at this point is quite small (just 8\%).
We find that PC = 32\% at $s_0 = 23$ GeV$^2$. Moreover, the $M_B$ dependence is the weakest at this point.
Accordingly, we fix the threshold value to be $s_0 = 23$ GeV$^2$ and choose $21$ GeV$^2\leq s_0\leq 25$ GeV$^2$ as our working region.

Finally, we change $\theta_1$ and redo the above processes. We show variations of $M_X$ with respect to $\theta_1$ in the right panel of Fig.~\ref{fig:mix1mb} when fixing $s_0= 23$ GeV$^2$ and choosing $M_B$ to satisfy CVG$=10\%$. We find that the $\theta_1$ dependence of the mass prediction is weak when $\theta_1 \leq -40^\circ$. Accordingly, we fix the mixing angle $\theta_1$ to be $-42^\circ$ and choose $\theta_1 = -42\pm5^\circ$ as our working region.

Altogether for the current $J_{\mu,3/2-}$, we fine-tune the mixing angle $\theta_1$ to be $-42^\circ$, and the working regions are found to be $21$ GeV$^2\leq s_0\leq 25$ GeV$^2$ and $2.59$ GeV$^2< M_B^2 < 3.19$ GeV$^2$. We assume the uncertainty of $\theta_1$ to be $-42\pm5^\circ$, and obtain the following numerical results:
\begin{eqnarray}
M_{3/2^-} &=& 4.40^{+0.17}_{-0.22} \mbox{ GeV} \, ,
\label{result:mix1}
\\ \nonumber f_{3/2^-} &=& (6.5^{+3.2}_{-2.9}) \times 10^{-4} \mbox{ GeV}^6 \, ,
\end{eqnarray}
where the central value corresponds to $\theta_1 = -42^\circ$, $s_0 = 23$ GeV$^2$ and $M_B^2=2.89$ GeV$^2$. The mass uncertainty is due to the mixing angle $\theta_1$, the Borel mass $M_B$, the threshold value $s_0$, the charm quark mass $m_c$, and various condensates~\cite{pdg,parameter1,parameter2,parameter3,parameter4,parameter5,parameter6,parameter7,parameter8}. We note that: a) when calculating the mass uncertainty due to the mixing angle $\theta_1$, we have fixed $s_0$ and $M_B$; and b) when plotting the mass variation as a function of $\theta_1$ as shown in the right panel of Fig.~\ref{fig:mix1mb}, we have fixed $s_0$ but choosing $M_B$ to satisfy CVG$=10\%$.
The above mass value is consistent with the experimental mass of the $P_c(4380)$~\cite{lhcb1}, and supports it to be a hidden-charm pentaquark having $J^P=3/2^-$.
The current $J_{\mu,3/2-}$ consists of $\xi_{36\mu}$ and $\psi_{9\mu\nu}$, suggesting that the $P_c(4380)$ may contain the $S$-wave $[\Lambda_c(1P)\bar D_1]$, $P$-wave $[\Lambda_c(1P)\bar D]$, $P$-wave $[\Lambda_c\bar D_1]$, $D$-wave $[\Lambda_c\bar D]$, and $S$-wave $[\Sigma_c \bar D^*]$ components, etc.

\begin{figure*}[!hbt]
\begin{center}
\scalebox{0.4}{\includegraphics{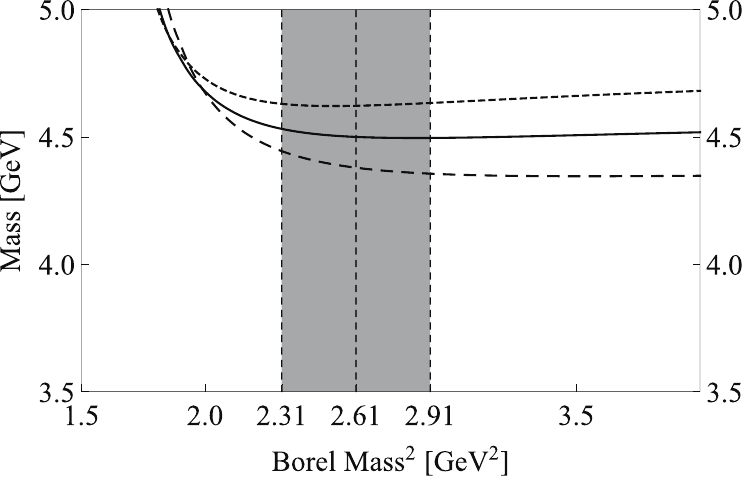}}
\scalebox{0.4}{\includegraphics{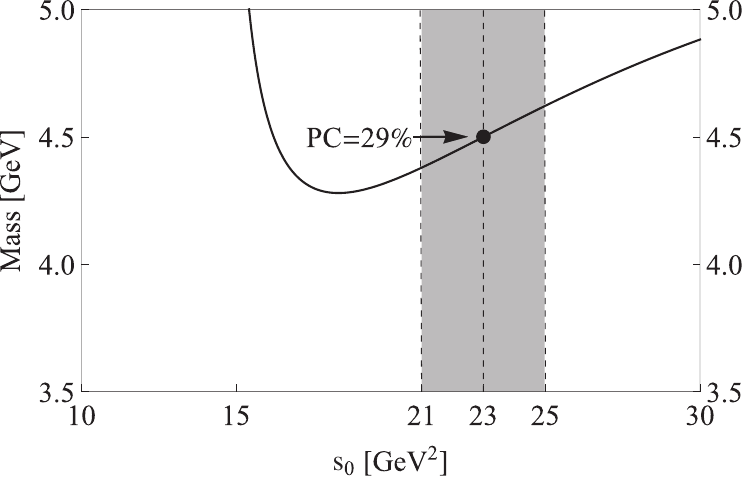}}
\scalebox{0.4}{\includegraphics{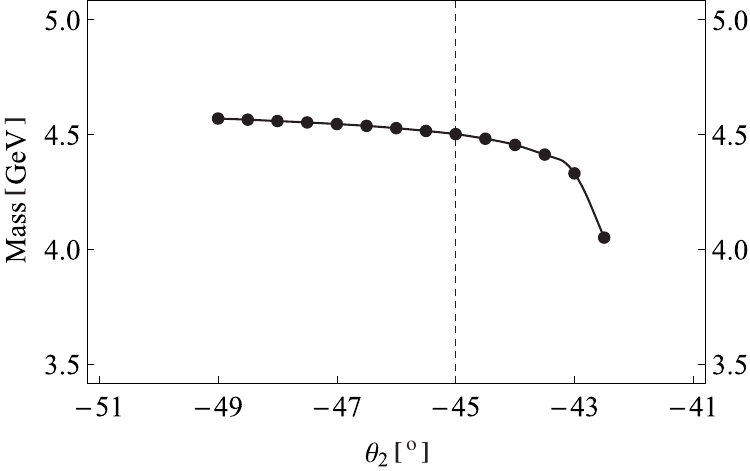}}
\caption{Variations of $M_{5/2^+}$ with respect to the Borel mass $M_B$ (left), the threshold value $s_0$ (middle) and the mixing angle $\theta_2$ (right),
calculated using the current $J_{\mu\nu,5/2+}$ of $J^P = 5/2^+$.
In the left figure, the long-dashed, solid and short-dashed curves are obtained with $\theta_2=-45^\circ$ and for $s_0 = 21$, $23$ and $25$ GeV$^2$, respectively.
In the middle figure, the curve is obtained with $\theta_2=-45^\circ$ and $M_B^2 = 2.61$ GeV$^2$.
In the right figure, the curve is obtained for $s_0 = 23$ GeV$^2$ and with $M_B$ satisfying CVG$=10\%$.}
\label{fig:mix2mb}
\end{center}
\end{figure*}

Similarly, we investigate the current $J_{\mu\nu,5/2+}$ of $J^P = 5/2^+$. We fine-tune the mixing angle $\theta_2$ to be $-45\pm5^\circ$, and the working regions are found to be $21$ GeV$^2\leq s_0\leq 25$ GeV$^2$ and $2.31$ GeV$^2< M_B^2 < 2.91$ GeV$^2$. We show variations of $M_X$ with respect to $M_B$, $s_0$, and $\theta_2$ in Fig.~\ref{fig:mix2mb}, and obtain the following numerical results:
\begin{eqnarray}
M_{5/2^+} &=& 4.50^{+0.26}_{-0.24} \mbox{ GeV} \, ,
\label{result:mix2}
\\ \nonumber f_{5/2^+} &=& (5.5^{+3.4}_{-2.4}) \times 10^{-4} \mbox{ GeV}^6 \, ,
\end{eqnarray}
where the central value corresponds to $\theta_2=-45^\circ$, $s_0 = 23$ GeV$^2$ and $M_B^2=2.61$ GeV$^2$.
The above mass value is consistent with the experimental mass of the $P_c(4450)$~\cite{lhcb1}, and supports it to be a hidden-charm pentaquark having $J^P=5/2^+$. The current $J_{\mu\nu,5/2+}$ consists of $\xi_{15\mu}$ and $\psi_{4\mu\nu}$, suggesting that the $P_c(4450)$ may contain the $S$-wave $[\Lambda_c(1P)\bar D^*]$, $P$-wave $[\Lambda_c\bar D^*]$, $S$-wave $[\Sigma_c^* \bar D_1]$, and $P$-wave $[\Sigma_c^* \bar D]$ components, etc.

\section{Other spin-parity assignments}
\label{sec:others}

In this section we follow the same approach to study the hidden-charm pentaquark states of $J^P = 3/2^+$ and $J^P = 5/2^-$. We find the following two currents
\begin{eqnarray}
J_{\mu,3/2+} &=& \cos\theta_3 \times \xi_{35\mu} + \sin\theta_3 \times \psi_{10\mu}
\label{def:mix3}
\\ \nonumber &=& \cos\theta_3 \times [\epsilon^{abc} (u^T_a C \gamma_\nu \gamma_5 d_b) \gamma_\nu \gamma_5 c_c] [\bar c_d \gamma_\mu u_d]
\\ \nonumber && + \sin\theta_3 \times [\epsilon^{abc} (u^T_a C \gamma_\nu u_b) \gamma_\nu \gamma_5 c_c] [\bar c_d \gamma_\mu \gamma_5 d_d] \, ,
\\
J_{\mu\nu,5/2-} &=& \cos\theta_4 \times \xi_{16\mu\nu} + \sin\theta_4 \times \psi_{3\mu\nu}
\label{def:mix4}
\\ \nonumber &=& \cos\theta_4 \times [\epsilon^{abc} (u^T_a C \gamma_\mu \gamma_5 d_b) c_c] [\bar c_d \gamma_\nu \gamma_5 u_d]
\\ \nonumber && + \sin\theta_4 \times [\epsilon^{abc} (u^T_a C \gamma_\mu u_b) c_c] [\bar c_d \gamma_\nu d_d]
\\ \nonumber && + \{ \mu \leftrightarrow \nu \} \, ,
\end{eqnarray}
which have structures similar to $J_{\mu,3/2-}$ and $J_{\mu\nu,5/2+}$, respectively.
The extracted spectral densities are also similar to previous ones:
\begin{eqnarray}
\rho_{3/2+,1}(s) &=& \rho^{pert}_{3/2-,1}(s) - \rho^{\qq}_{3/2-,1}(s) + \rho^{\GGa}_{3/2-,1}(s)
\\ \nonumber && - \rho^{\qq^2}_{3/2-,1}(s) + \rho^{\qGqa}_{3/2-,1}(s) + \rho^{\qq\qGqa}_{3/2-,1}(s) \, ,
\\ \rho_{5/2-,1}(s) &=& \rho^{pert}_{5/2+,1}(s) - \rho^{\qq}_{5/2+,1}(s) + \rho^{\GGa}_{5/2+,1}(s)
\\ \nonumber && - \rho^{\qq^2}_{5/2+,1}(s) + \rho^{\qGqa}_{5/2+,1}(s) + \rho^{\qq\qGqa}_{5/2+,1}(s) \, ,
\end{eqnarray}
where $\rho^{pert}_{3/2-,1}(s)$, $\rho^{pert}_{5/2+,1}(s)$ and others have been given in Eqs.~(\ref{ope:mix1}) and (\ref{ope:mix3}).

\begin{figure*}[hbt]
\begin{center}
\scalebox{0.6}{\includegraphics{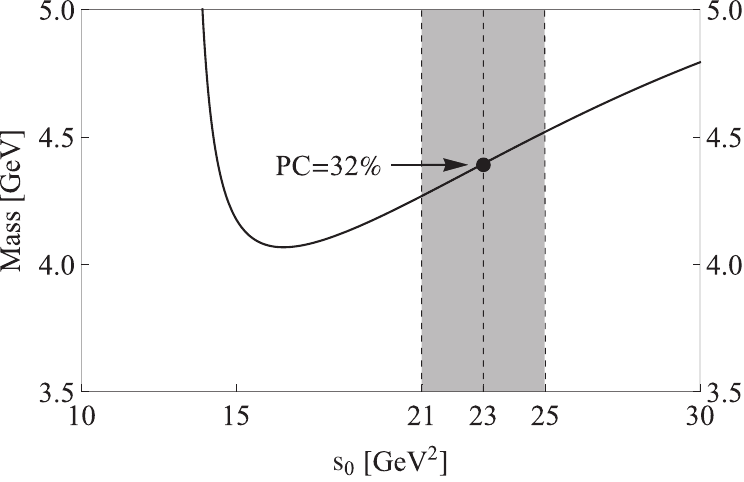}}
\scalebox{0.6}{\includegraphics{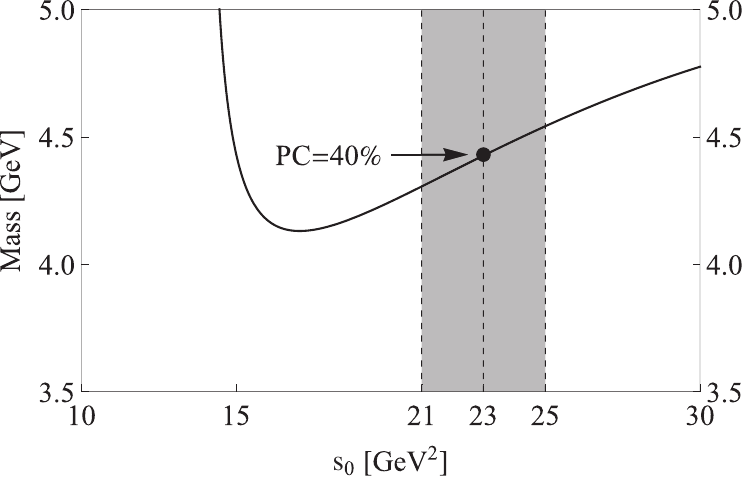}}
\caption{Variations of $M_{3/2^+}$ (left) and $M_{5/2^-}$ (right) with respect to the threshold value $s_0$,
calculated using the current $J_{\mu,3/2+}$ with $\theta_3=-42^\circ$ and $J_{\mu\nu,5/2-}$ with $\theta_4=-45^\circ$, respectively.}
\label{fig:mix34}
\end{center}
\end{figure*}

Firstly, we study the current $J_{\mu,3/2+}$ of $J^P = 3/2+$. With the same mixing angle as $\theta_1$, i.e., $\theta_3 = \theta_1 = -42\pm5^\circ$, the working regions are found to be $21$ GeV$^2\leq s_0\leq 25$ GeV$^2$ and $2.58$ GeV$^2< M_B^2 < 3.18$ GeV$^2$. We show variations of $M_X$ with respect to $s_0$ in the left panel of Fig.~\ref{fig:mix34} with $\theta_3=-42^\circ$, where the mass is extracted to be
\begin{eqnarray}
M_{3/2^+} &=& 4.40^{+0.14}_{-0.16} \mbox{ GeV} \, .
\label{result:mix3}
\end{eqnarray}
Then we study the current $J_{\mu\nu,5/2-}$ of $J^P = 5/2-$. With the same mixing angle as $\theta_2$, i.e., $\theta_4= \theta_2 = -45\pm5^\circ$, the working regions are found to be $21$ GeV$^2\leq s_0\leq 25$ GeV$^2$ and $2.20$ GeV$^2< M_B^2 < 2.80$ GeV$^2$.  We show variations of $M_X$ with respect to $s_0$ in the right panel of Fig.~\ref{fig:mix34} with $\theta_4=-45^\circ$, where the mass is extracted to be
\begin{eqnarray}
M_{5/2^-} &=& 4.43^{+0.26}_{-0.28} \mbox{ GeV} \, .
\label{result:mix4}
\end{eqnarray}
The above two values are both consistent with the experimental masses of the $P_c(4380)$ and $P_c(4450)$~\cite{lhcb1}, suggesting that their spin-parity assignments can be different from $J^P=3/2^-$ and $5/2^+$, and further theoretical and experimental efforts are required to clarify their properties.

\section{Results and discussions}
\label{sec:summary}

In this paper we use the method of QCD sum rules to study the hidden-charm pentaquark states $P_c(4380)$ and $P_c(4450)$.
We achieve better QCD sum rule results by requiring the pole contribution to be larger than or around 30\% to insure the one-pole parametrization to be valid, which criterion is more strict than that used in our previous studies~\cite{chen1,chen2}. We find two mixing currents, $J_{\mu,3/2-}$ of $J^P = 3/2^-$ and $J_{\mu\nu,5/2+}$ of $J^P = 5/2^+$. We use them to perform sum rule analyses, and the masses are extracted to be
\begin{eqnarray*}
M_{3/2^-} &=& 4.40^{+0.17}_{-0.22} \mbox{ GeV} \, ,
\\ M_{5/2^+} &=& 4.50^{+0.26}_{-0.23} \mbox{ GeV} \, .
\end{eqnarray*}
These values are consistent with the experimental masses of the $P_c(4380)$ and $P_c(4450)$, suggesting that they can be identified as hidden-charm pentaquark states composed of anti-charmed mesons and charmed baryons: the $P_c(4380)$ has $J^P=3/2^-$ and may contain the $S$-wave $[\Lambda_c(1P)\bar D_1]$, $P$-wave $[\Lambda_c(1P)\bar D]$, $P$-wave $[\Lambda_c\bar D_1]$, $D$-wave $[\Lambda_c\bar D]$, and $S$-wave $[\Sigma_c \bar D^*]$ components, etc; the $P_c(4450)$ has $J^P=5/2^+$ and may contain the $S$-wave $[\Lambda_c(1P)\bar D^*]$, $P$-wave $[\Lambda_c\bar D^*]$, $S$-wave $[\Sigma_c^* \bar D_1]$, and $P$-wave $[\Sigma_c^* \bar D]$ components, etc.

We follow the same approach to study the hidden-charm pentaquark states of $J^P = 3/2^+$ and $J^P = 5/2^-$, and extract their masses to be
\begin{eqnarray*}
M_{3/2^+} &=& 4.40^{+0.14}_{-0.16} \mbox{ GeV} \, ,
\\ M_{5/2^-} &=& 4.43^{+0.26}_{-0.28} \mbox{ GeV} \, .
\end{eqnarray*}
These values are also consistent with the experimental masses of the $P_c(4380)$ and $P_c(4450)$~\cite{lhcb1}, suggesting that there still exist other possible spin-parity assignments for them, which needs to be clarified in further theoretical and experimental studies.

\begin{figure*}[hbt]
\begin{center}
\scalebox{0.6}{\includegraphics{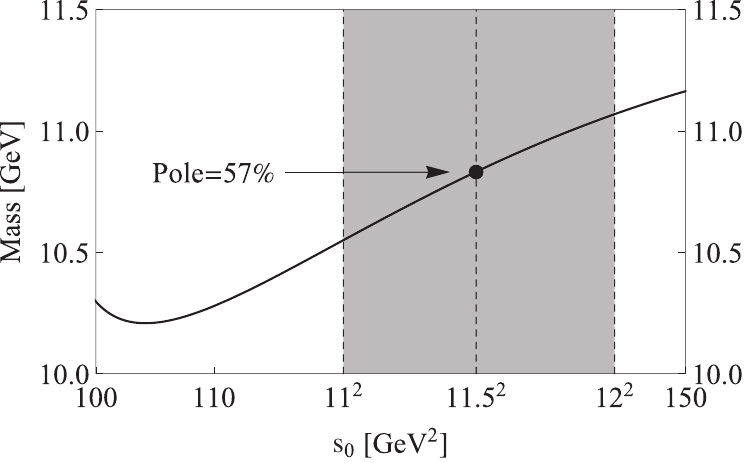}}
\scalebox{0.6}{\includegraphics{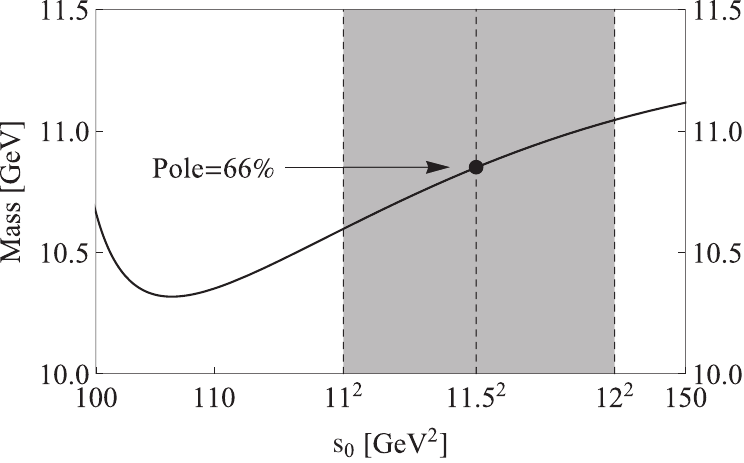}}
\caption{Variations of $M_{P_b(3/2^-)}$ (left) and $M_{P_b(5/2^+)}$ (right) with respect to the threshold value $s_0$,
calculated using the current $J^{b \bar b u u d}_{\mu,3/2-}$ with $\theta_1=-42^\circ$ and $J^{b \bar b u u d}_{\mu\nu,5/2+}$ with $\theta_2=-45^\circ$, respectively.}
\label{fig:bottom}
\end{center}
\end{figure*}

We have also investigated the bottom partners of the $P_c(4380)$ and $P_c(4450)$, i.e., the hidden-bottom pentaquark states ($b \bar b u u d$) of $J^P = 3/2^-$ and $J^P = 5/2^+$. As shown in Fig.~\ref{fig:bottom}, their masses are extracted to be
\begin{eqnarray}
M_{P_b(3/2^-)} &=& 10.83^{+0.26}_{-0.29} \mbox{ GeV} \, ,
\\ \nonumber M_{P_b(5/2^+)} &=& 10.85^{+0.24}_{-0.27} \mbox{ GeV} \, .
\end{eqnarray}
We propose to search for them in the future LHCb and BelleII experiments.

To end this paper, we note that there are quite some systematical uncertainties which are not considered in the present study, such as the vacuum saturation for higher dimensional operators which is used when calculating the OPE serious. Moreover, in this paper we have used the running charm and bottom quark masses in the $\overline{MS}$ scheme, while sometimes their pole masses are used instead. Take the current $J_{\mu,3/2-}$ as an example: a) if we use $\bra 0 | \bar q q \bar q q |0\ket = ( 0.8 \sim 1.2 ) \times \bra 0 | \bar q q |0\ket \bra 0|\bar q q |0\ket$, we would obtain $M_{3/2^-} = 4.34$ GeV $\sim 4.48$ GeV (other uncertainties are not included);  b) if we use the pole charm mass $m_c = 1.67$~GeV~\cite{pdg}, we would have to shift the mixing angle to be around $\theta_1 = -38^\circ$ in order to arrive at the similar mass $M_{3/2^-} = 4.38$ GeV. Combining the previous uncertainties in Eqs.~(\ref{result:mix1}), (\ref{result:mix2}), (\ref{result:mix3}) and (\ref{result:mix4}), we obtain the following result for the mixing current $J_{\mu,3/2-}$ of $J^P = 3/2^-$
\begin{eqnarray*}
M_{3/2^-} &=& 4.40^{+0.19}_{-0.23} \mbox{ GeV} \, .
\end{eqnarray*}
Similarly, we obtain the following results for the other three mixing currents, $J_{\mu\nu,5/2+}$ of $J^P = 5/2^+$, $J_{\mu,3/2+}$ of $J^P = 3/2^+$, and $J_{\mu\nu,5/2-}$ of $J^P = 5/2^-$:
\begin{eqnarray*}
M_{5/2^+} &=& 4.50^{+0.28}_{-0.27} \mbox{ GeV} \, ,
\\ M_{3/2^+} &=& 4.40^{+0.16}_{-0.17} \mbox{ GeV} \, ,
\\ M_{5/2^-} &=& 4.43^{+0.27}_{-0.29} \mbox{ GeV} \, .
\end{eqnarray*}
The above (systematical) uncertainties are significant, suggesting that we still know little about exotic hadrons, and further experimental and theoretical studies are needed in order to well understand them.

\acknowledgments{We thank Professor Nikolai Kochelev for helpful discussions.}

\end{multicols}

\vspace{10mm}

\vspace{-1mm}
\centerline{\rule{80mm}{0.1pt}}
\vspace{2mm}

\begin{multicols}{2}

\end{multicols}

\clearpage

\end{document}